\documentclass[aps,pra,amsmath, amssymb, english, twocolumn,reprint,floatfix, superscriptaddress]{revtex4-1}

\usepackage{bbm}
\usepackage{comment}

\usepackage{tikz}
\usepackage{pgfplots}
\usepackage{microtype}
\usepackage{tkz-euclide}
\usetkzobj{all}
\usepackage{tkz-graph}
\usepackage{epstopdf}
\usepackage{epsfig} 
\usepackage[applemac]{inputenx}     
\usepackage{amsmath}

\usepackage{hyperref}
\hypersetup{
    colorlinks,
    citecolor=blue,
    filecolor=blue,
    linkcolor=blue,
    urlcolor=blue
}

\newcommand{\bR}{\mathbf{R}}

\newcommand{\cc}{c^{\ }}
\newcommand{\bq}{\mathbf{q}}

\newcommand{\bk}{\mathbf{k}}

\newcommand{\heff}{\hat{H}_{\mbox{\footnotesize \emph{eff}}} }

\newcommand{\cd}{c^\dag}

\newcommand{\av}[1]{\left <#1\right>}

\allowdisplaybreaks

\graphicspath{{./Images/}}
\begin{document}

\title{Selective insulators and anomalous responses in correlated fermions with synthetic extra dimensions}

\author{Lorenzo Del Re} 
\affiliation{International School for
  Advanced Studies (SISSA), Via Bonomea
  265, I-34136 Trieste, Italy}
\affiliation{Institute for Solid State Physics, TU Wien, 1040 Vienna, Austria}
\author{Massimo Capone} 
\affiliation{International School for
  Advanced Studies (SISSA), Via Bonomea
  265, I-34136 Trieste, Italy} 
\affiliation{CNR-IOM Democritos, Via Bonomea
  265, I-34136 Trieste, Italy}

\date{\today} 


\begin{abstract}
We study a three-component fermionic fluid in an optical lattice in a regime of intermediate-to-strong interactions allowing for Raman processes connecting the different components, similar to those used to create artificial gauge fields (AGF). Using Dynamical Mean-Field Theory we show that the combined effect of interactions and  AGFs induces a variety of anomalous phases in which different components of the fermionic fluid display qualitative differences, i.e., the physics is flavor-selective. Remarkably, the different components can display huge differences in the correlation effects, measured by their effective masses and non-monotonic behavior of their occupation number as a function of the chemical potential, signaling a sort of selective instability of the overall stable quantum fluid.
\end{abstract}
\maketitle
Gases of ultracold Ytterbium atoms ($^{173}$Yb) have emerged as an incredibly rich toolbox for the the quantum simulations of  multi-component fermions.  
$^{173}$Yb atoms feature a large nuclear spin $I = 5/2$ with six possible spin flavors. The interaction between atoms are  independent on the nuclear spin index. This leads to the possibility of the direct quantum simulation of $SU(N)$-symmetric Hubbard models \cite{gorshkov2010two,pagano2014one,
zhang2015orbital,PhysRevLett.115.265301,PhysRevLett.115.265302,PhysRevLett.113.120402,scazza2014observation} once the gas is confined in a sufficiently deep optical lattice.

New physical regimes can be explored explicitly breaking the $SU(N)$ symmetry by exploiting atom-light interaction processes that couple the internal degrees of freedom of the atoms. Such a coupling can be realized through Raman transitions between atomic levels with different nuclear spin, which have been used to design the so-called Artificial Gauge Fields (AGF) \cite{dalibard2011colloquium, goldman2014light, tagliacozzo2013simulation,celi2014synthetic,0034-4885-78-2-026001,mancini2015observation,PhysRevLett.115.095302,barbarino2015magnetic}. In this way one can mimic tunable static electric and magnetic fields, as well as spin-orbit coupling, despite the neutrality of the atoms\cite{PhysRevLett.109.095301,PhysRevLett.111.225301,song2016spin,huang2016experimental}. 
The identification of Raman processes with a gauge field is based on interpreting the spin degree of freedom as a discrete artificial spatial dimension which adds to the dimensionality of  the optical lattice \cite{PhysRevLett.108.133001,celi2014synthetic,mancini2015observation}. In this language the Raman transitions between atomic levels with different nuclear spin, correspond to a quantum tunnelling or hopping between different sites of the synthetic dimension. These hoppings can have a phase\cite{celi2014synthetic} which is used to mimic a Peierls phase.
This scheme has been applied to cold atoms of $^{173}$Yb confined in a one dimensional optical array for the detection of chiral edge states localized at the edges  of  the synthetic dimension \cite{mancini2015observation}.

 In the present work, we take a slightly different point of view and we focus on the regime of large interactions, which can be realized by means of a deep optical lattice. In particular we target the possibility to induce via the combined effect of strong interactions and of the AGF, what we call a {\it{selective}} behavior, namely a regime where different "parts" of the fermionic fluid have qualitative different properties induced by the interactions.  The possibility of selective phases of matter in multicomponent systems in the presence of strong correlations is emerging as a new paradigm in solid-state physics. The epitome of these phenomena is the orbital-selective Mott insulator\cite{Vojta2010}, a state where some orbital degrees of freedom are localized by interactions (as in a Mott insulator), while others remain metallic. This regime, whose very existence was questioned until recently, is particularly interesting because the scattering of the mobile carriers with the localized partners can lead to anomalous properties and a breakdown of the standard metallic behavior described by Fermi-liquid theory.
 
These concepts have been proposed in various models\cite{PhysRevLett.91.226401,PhysRevLett.92.216402,PhysRevB.72.205124,PhysRevB.72.205126,PhysRevLett.99.126405,PhysRevLett.102.126401}, but they found a new life after the discovery of iron-based superconductors\cite{Hosono}, where the evidence of orbital-selective correlations\cite{PhysRevLett.112.177001,PhysRevB.94.205113,PhysRevB.96.045133} and even an orbital-selective Mott transition\cite{PhysRevLett.110.067003,PhysRevLett.110.146402} is now very strong, and an orbital-selective superfluid pairing has been experimentally reported for FeSe\cite{Sprau75}. However, the physics of iron-based superconductors is complicated by hybridization between orbitals and various material-dependent effects. Therefore the experimental realization of similar selective phases with ultracold gases and optical lattices can help to shed light on these novel phenomena and their relation with anomalous metallic properties and superfluidity.

In order to focus on the very existence of selective phases, we consider one of the simplest set-ups, a three-component gas obtained by selecting only three spin flavors and the hopping amplitudes in the synthetic dimension assumed to be real numbers. 
The latter choice would correspond to zero magnetic flux in the language of gauge fields, but it is sufficient to highlight the selective physics we address here. Despite the looser connection with gauge fields, in the following we will define the Raman-induced fields as AGF to establish a connection with other studies. Even if experimental studies have so far focused on the case of finite-momentum two-photon processes, which lead to complex tunneling amplitudes, the present case can be realized. From a theoretical point of view, this allows to address the effects of the interplay between AGF and Mott physics in its simplest realization.

The analysis we report in this paper demonstrates the existence of various remarkable selective phenomena driven by the value of the AGF amplitudes as a function of the overall density of atoms. In a basis which diagonalizes the AGF amplitudes, we find that the effective masses of two components increase dramatically when a metal-insulator transition is approached, while the third component approaches its non-interacting value. In the case where the amplitudes are chosen in order to leave two degenerate combinations we find a transition towards an exotic mixture between a Mott insulator and a totally polarized state.  In both cases we observe anomalous dependencies of flavor-resolved densities.

\section{Model and Methods}
The Hamiltonian of our multicomponent system in a grandcanonical ensembles reads
\begin{eqnarray}\label{Hub:Ham:1}
\hat{H} &=& -t\sum_{\left<\bR\bR^\prime\right>\sigma}c^{\dag }_{\bR \sigma}\,c^{\ }_{\bR^\prime \sigma } + \sum_{\bR,\sigma\sigma^\prime} \,c^{\dag}_{\bR \sigma}\tau_{\sigma\sigma^\prime}\,c^{\ }_{\bR \sigma^\prime} +
 \nonumber \\ 
 &U&\sum_{\bR,\,\sigma< \sigma^\prime} \left(\hat{n}^{\ }_{\bR \sigma} - \frac{1}{2}\right)
\left(\hat{n}^{\ }_{\bR \sigma^\prime}-\frac{1}{2}\right) -\mu\sum_{\bR \sigma} \hat{n}^{\ }_{\bR \sigma}
\nonumber \\
\end{eqnarray}
where $c^{\ }_{\bR \sigma}$ is the destruction operator of a Fermion with spin $\sigma=1,2, \ldots, N$ on lattice site $\bR$, $\hat{n}^{\  }_{\bR \sigma} \equiv c^{\dag }_{\bR \sigma} c^{\ }_{\bR \sigma}$, $\mu$ is the chemical potential which fixes the average particle density, and $\tau_{\sigma\sigma^\prime}$ is the  $N \times N$ AGF matrix that couples the internal degrees of freedom locally. Here we limit to $N=3$ and to real amplitudes. Therefore, $\tau_{\sigma\sigma^\prime}$ is a real symmetric matrix with all diagonal elements set to zero. 

\begin{figure}[hbtp!]
%

\includegraphics[width = 0.5\textwidth]{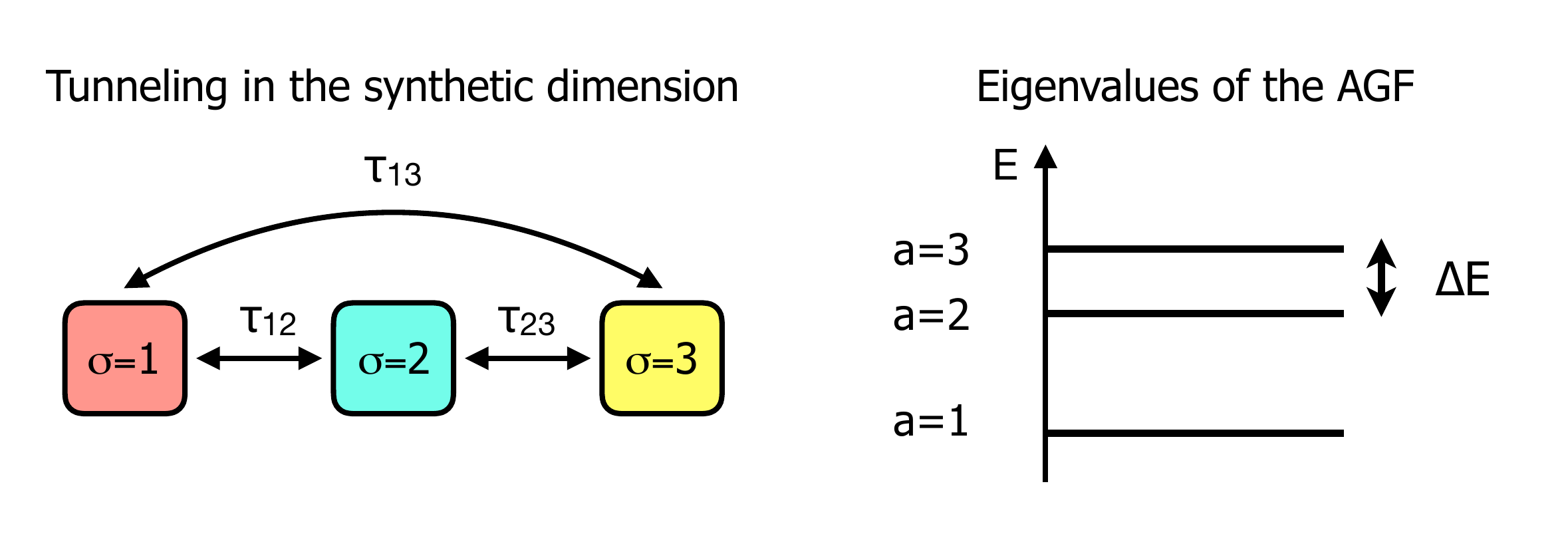}
\caption{Schematic representation of the hopping processes along the synthetic dimension and their energy levels in the $SU(3)$ case.}

\label{PBC_OBC_synthetic}
\end{figure}

If we interpret the internal spin index as running along a synthetic spatial dimension, $\tau_{\sigma\sigma^\prime}$ simply measures the tunneling amplitude between "sites" $\sigma$ and $\sigma^\prime$, as depicted in Fig. \ref{PBC_OBC_synthetic}.  These hopping processes can be experimentally simulated by means of Raman transitions between the hyperfine levels of $^{173}$Yb \cite{mancini2015observation}. So far, the case of "nearest-neighbor" hopping $\tau_{13} = 0$ has been experimentally considered, but also next-neighbor processes can be induced.  

The AGF clearly split the $SU(3)$ degeneracy between the local spin levels. 
In a generic case where $\tau_{\sigma\sigma^\prime}$ have no particular symmetry the degeneracy is completely lifted, the secular equation is given by $\lambda^3 +p \lambda + q = 0$, where $p = -(\tau_{12}^2 + \tau_{23}^2+\tau_{13}^2)$ and $q = - 2\,\tau_{12}\tau_{23}\tau_{13}$ and we have three different eigenvalues $\lambda_a = 2\sqrt{-\frac{p}{3}}\cos\left(\frac{2\pi (a-1)+\theta}{3}\right)$, with $\theta = \mbox{Arg}\left(-q/2 + i\sqrt{-\Delta}\right)$, $\Delta = \frac{q^2}{4}+\frac{p^3}{27} \leq 0$ and $a = 1, 2,...,N$.
When 
$\tau_{13} = 0$, i.e. in the case with no next-nearest-neighbors hopping in the synthetic dimension, there is always a vanishing eigenvalue and the energy levels are equally spaced, namely $\{\lambda\} = \{-\sqrt{\tau_{12}^2 + \tau_{23}^2},0,\sqrt{\tau_{12}^2 + \tau_{23}^2}\}$.
For choices of parameters such that
$\Delta = 0$ (e.g. when $\tau_{12} = \tau_{23} = \tau_{13}$) two eigenvalues are degenerate and a residual $SU(2)$ symmetry survives. 
Since the interaction is invariant under any canonical transformation in the spin space, it is very useful to work in the basis which diagonalizes the AGF matrix, where the new fermion destruction operator reads $\widetilde{c}^{\ }_{a} = \sum_{\sigma}U_{a\sigma}\,c^{\ }_{\sigma}$, with $\sum_{\sigma \sigma^\prime}U_{a\sigma}\tau_{\sigma \sigma^\prime}U^\dag_{\sigma^\prime b} \equiv \lambda_{a}\delta_{ab}$.
In this new basis,  the effect of the AGF is reflected in different local energy levels $\lambda_a$ which resemble a crystal-field splitting of a degenerate level in a solid or, if we prefer to maintain a language connected to magnetism, a kind of magnetic field which induces some splittings between the different eigenlevels. In the following we will always refer to this diagonal basis and we shall drop the tilde to lighten the notation.
The use of the basis has several advantages, besides the formal simplicity. This basis is indeed the ideal one to characterize orbital-selective phenomena, and it simplifies the theoretical treatment because it does not require the evaluation of Green's functions and self-energy non-diagonal in the spin index. 

We also notice that the same modelization we use for the diagonal basis could be obtained for other experimental set-ups where the occupations of some orbitals are different, starting from the simplest realization of a system where different number of atoms are loaded in the various spin states. 


We addressed the hamiltonian in Eq. (\ref{Hub:Ham:1}) using Dynamical Mean Field Theory (DMFT) \cite{georges1996dynamical}, an approach which treats exactly local quantum correlations by mapping the lattice model onto an effective theory where a single interacting site is embedded in a effective non interacting medium to be determined self-consistently or, in other words, a self-consistent impurity model. The method becomes exact in the limit of infinite spatial coordination or connectivity.

Within DMFT we assume that any site is equivalent for the description of the system and we pick an arbitrary site "0" and we replace its interaction with the rest of the lattice with an effective bath which has to be determined self-consistently. This defines an effective impurity model where site 0 is the only interacting site (the impurity) and some auxiliary non-interacting fermions are introduced to describe the bath. In our case this reads
\begin{eqnarray}\label{imp:ham:AGF}
 \hat{H}_{\mbox{\scriptsize \emph{eff}}} &=& \sum_{\ell a} \epsilon_{\ell a}\cd_{\ell a}c^{\ }_{\ell a} + \sum_{\ell a}\left(V_{\ell a}\cd_{\ell a}\cc_{0a} + \mbox{h.c.}\right)\nonumber \\
&+& U\sum_{a<b}\hat{n}_{0a} \hat{n}_{0b} + \sum_a(\lambda_a - \mu)\hat{n}_{0a}, 
 \end{eqnarray}
 where  $\hat{n}_{0a} = \cd_{0a} \cc_{0a}$ is the occupation operator for the different spin states of the impurity site, $\cd_{\ell a}$ is the creation operator of a fermion in a level of the non-interacting bath, $\epsilon_{\ell a}$ are the energy levels of the effective bath and $V_{\ell a}$ are  hybridization amplitudes between the interacting impurity and the effective bath, $\mu$ is the chemical potential. 
It is worth to remind that the basic control parameters are the Hubbard repulsion $U$ and the AGF splittings $\lambda_a$, and a crucial physical quantity is the lattice filling $n = N_f/N_s$, where $N_f$ is the number of fermions and $N_s$ is the number of sites, which is fixed by the chemical potential. In fact, all other parameters in eq.(\ref{imp:ham:AGF}) have to be determined self-consistently.

 The mapping between the lattice model and the effective local theory is realized by a self-consistency condition which requires that the Green's function of the correlated site in the effective model $G_a(\tau) = -T_\tau\av{\cc_{0a}(\tau)\cd_{0a}(0)}_{\heff}$ coincides with the local component of the lattice Green's function. This condition depends on the original lattice, which enters only through the bare density of states. 
 
In this paper we work on a Bethe lattice with a  semicircular density of states (DOS) $g(\epsilon)=\frac{2}{\pi D}\sqrt{D^2-\epsilon^2}$, which provides accurate results for three-dimensional lattices with a simplified self-consistency condition

\begin{eqnarray}\label{DMFT:eqs:AGF}
\mathcal{G}^{-1}_a(i\nu) &=& i\nu +\mu -\lambda_a -\frac{D^2}{4}G_a(i\nu),
\end{eqnarray}
where $\mathcal{G}_a^{-1}$ is the non-interacting Green's function of the effective model, which plays the role of a dynamical effective field and reads
\begin{equation}
\mathcal{G}_a^{-1}(i\nu) = i\nu +\mu -\lambda_a-\sum_\ell \frac{|V_{\ell a}|^2}{i\nu-\epsilon_{\ell a}}.
\end{equation}

In our implementation the effective impurity model is solved using a Lanczos exact diagonalization solver. Within this scheme\cite{CaffarelKrauth,CaponeED} the effective bath must be represented as a finite-size matrix, which is obtained considering a finite number of bath levels. It has been shown that even very small number of levels are sufficient to provide very accurate results for thermodynamic observables and here we consider $18$ spin-orbital levels.

\section{Linear response to the artificial Gauge fields}

Before addressing the interplay of the AGF with strong local interactions, we present a linear-response analysis of the effects of the AGF in the diagonal basis, where we can make a direct analogy with magnetism in the standard SU(2) case.

On very general grounds an increase of the magnetic response is expected as the interaction increases. When we approach a Mott transition the fermions are progressively localized and they retain only the spin degree of freedom becoming more and more susceptible to an external magnetic field. For this reason the enhancement of the magnetic response is particularly strong at half-filling, but it survives also at finite doping. Within DMFT it  has been proven that the uniform magnetic susceptibility increases, but remains finite at the Mott localization transition, in contrast with the divergent local susceptibility reflecting the formation of local magnetic moments\cite{georges1996dynamical}. Here we shall give a generalization of this findings for the $SU(N)$ case based on the discussion of  Ref. \cite{georges1996dynamical}.

In the diagonal basis we can write the AGF Hamiltonian as $\hat{H}_{AGF} = -\sum_\bR\sum_{a = 1}^{N} \lambda_{\bR a} \hat{n}_{\bR a}$, which we now treat as a perturbation of the Hubbard Hamiltonian, where $\lambda_{\bR\alpha}$ are the on-site energies of the different fermionic species that satisfy the following relation $\sum_a \lambda_{\bR a} = 0$. In order to discuss general magnetic responses, we maintain a dependence of the AGF on the site $\bR$ even though in our DMFT calculations we will assume site-independent fields.

For every site we have $N-1$ independent variables that we call $h_{\bR\alpha}$ with $\alpha = 1, 2, ... ,N-1$. If we set $\lambda_{\bR a} = h_{\bR \alpha}$ for $a < N$ and $\lambda_{\bR N} = -\sum_{\alpha = 1}^{N-1}h_{\bR \alpha}$, we can write the AGF term as $\hat{H}_{AGF} = -\sum_{\alpha = 1}^{N-1}\sum_\bR h^\alpha_\bR \hat{S}^{\alpha}_\bR$, where $\hat{S}^{\alpha}_\bR \equiv \hat{n}_{\bR \alpha}-\hat{n}_{\bR N}$.

In the $N=2$ case we have only one independent variable, that is the magnetic field $h$ that multiplies the z- component of the magnetization $\hat{S}^{z}$ operator and we define a magnetic susceptibility $\chi(\bR-\bR^\prime) \equiv \left.\frac{\partial \langle\hat{S}_\bR\rangle}{\partial h^{\bR^\prime}}\right|_{h = 0}$ whose Fourier transform describes the tendency towards magnetic ordering of the system.

In the $SU(N)$ case we can generalize the definition as $\chi^{\alpha}_\beta(\bR-\bR^\prime) \equiv \left.\frac{\partial \langle\hat{S}^\alpha_\bR\rangle}{\partial h^\beta_{\bR^\prime}}\right|_{h = 0}$ and the Fourier transform $\chi^\alpha_\beta(\bR-\bR^\prime) = \frac{1}{V}\sum_{\bq}e^{i\mathbf{q}\cdot(\bR-\bR^\prime)}\chi^\alpha_\beta(\mathbf{q}) $.

We can then write a generalized Bethe-Salpeter equation, which we derive in Appendix \ref{Der:BS} and reads:
\begin{equation}\label{Bethe:Salpeter}
\left[\chi^\alpha_\beta\right]^{-1} = \left[\chi_0^{-1} + T\, \Gamma_{A} \right]/(1+ \delta_{\alpha\beta}),
\end{equation}
where $[\chi^\alpha_\beta]_{\nu\nu^\prime}$ enters as a generalized susceptibility from which the thermodynamic susceptibility is obtained by summing over the internal frequencies $\chi^\alpha_\beta(\mathbf{q}) = T^2\,\sum_{\nu\nu^\prime}[\chi^\alpha_\beta]_{\nu\nu^\prime}$. The whole equation must therefore be interpreted as a matrix identity in the space of Matsubara frequencies $\nu$ and $\nu'$ .   
$\Gamma_A$ is the antisymmetric component of the irreducible vertex function in the particle-hole channel, and $[\chi_0(\bq)]_{\nu \nu^\prime} \equiv- \frac{1}{VT}\sum_{\bk} G(\bk,\nu)G(\bk + \bq,\nu)\,\delta_{\nu\nu^\prime} $ is the bare susceptibility computed through the knowledge of the Green's function $G(\bk,\nu)$. 

Eq.(\ref{Bethe:Salpeter}), is very similar to the one obtained for the $SU(2)$-symmetric case. The only difference consists in the factor $1 + \delta_{\alpha\beta}$, that arises from 
the increased number of possible permutations of spin indices.

Within the Random Phase Approximation $\Gamma_A \sim -U\,T$ and it does not depend on the spin nor on the total number of species. 
Therefore all the components in the spin basis are identical to the single susceptibility of the $SU(2)$ model\cite{cazalilla2009ultracold} except for the $1+\delta_{\alpha\beta}$ factor in eq.(\ref{Bethe:Salpeter}), that becomes irrelevant for the study of broken symmetry phases.

In the limit of infinite coordination only the local component of the vertex enters in the response functions\cite{georges1996dynamical}. Therefore we can replace $\Gamma_A$ with  $\Gamma_A^{loc}$, the vertex function, which can be obtained from the Bethe-Salpeter equation relative to the effective impurity model in eq.(\ref{imp:ham:AGF}).

The latter is very similar to eq.(\ref{Bethe:Salpeter}) with the only difference that $\chi^{\alpha\,loc}_\beta$  and $\chi^{loc}_0$ do not depend on the momenta. From this equation we can obtain $\Gamma_A^{loc}$ and rewrite the Bethe-Salpeter equation for the lattice model in eq.(\ref{Hub:Ham:1}) as $ [\chi^{\alpha}_\beta(\mathbf{q})]^{-1} = \left([\chi_0(\mathbf{q})]^{-1} -[\chi_0^{loc}]^{-1}\right)/(1+\delta_{\alpha\beta})+ [\chi^{\alpha\,loc}_{\beta}]^{-1}$. 

This expression can be further simplified in the infinite-coordination Bethe lattice, where $\left([\chi_0(\mathbf{0})]^{-1} -[\chi_0^{loc}]^{-1}\right)_{\nu\nu^\prime} = D^2/4\delta_{\nu\nu^\prime}$ \ \cite{georges1996dynamical}, so that:
\begin{equation}\label{Bethe:final}
[\chi^\alpha_\beta(\mathbf{0}) ]^{-1} = \delta_{\nu\nu^\prime}D^2/[4(1 + \delta_{\alpha\beta})] + [\chi^{\alpha\, loc}_{\beta}]^{-1}.
\end{equation}

This relation clearly shows that, even though the local susceptibility diverges, as it happens approaching the Mott transition, the uniform susceptibility remains finite with a cut-off related to the bandwidth $\chi^{\alpha}_\beta = T^2\sum_{\nu\nu^\prime}[D^2/[4(1+\delta_{\alpha\beta})]\delta_{\nu\nu^\prime} + [\chi^{\alpha\, loc}_{\beta}]^{-1}]^{-1}$. 

Nevertheless, even though spontaneous breaking of the spin degeneracy is ruled out, we can expect a strong response to a finite field, what is usually referred to as \emph{metamagnetism}. Based on this simple perturbative finding, we expect the response of the system to the AGF to be particularly sensitive to the degree of correlation and to the proximity to a Mott transition. This implies that the bare differences in local energy will be reflected in enhanced polarizations which, as we shall see in the DMFT analysis, can strongly influence the physics of the system.

\section{Dynamical Mean-Field Theory Results}

The $SU(3)$ symmetric model in absence of AGF has been studied using DMFT in\cite{gorelik2009mott}, focusing on the Mott metal-insulator transition which generalizes the results for the SU(2) Hubbard model. Mott localization occurs upon increasing the Coulomb interaction when the number of carriers per lattice site is integer. In this case, when $U$ is the largest energy scale, the fermions are localized because of the high energy cost associated with charge fluctuations.
For the SU(3) model, a Mott transtion is therefore found for the two possible integer fillings $n=1$ and $n=2$ (obviously $n=3$ describes a trivial completely filled system). Interestingly, a half-filled lattice ($n=1.5$) does not correspond to an integer number of fermions per site, so a Mott transition is not possible, in contrast with the standard results for SU(2) models and with models with an even number of components.

In this section, we will not repeat the same analysis of the authors in\cite{gorelik2009mott}, but we will focus on the response of the system to the artificial gauge fields. Since, as mentioned above, we will work in the basis where the AGF are diagonal and their effect is similar to a magnetic field, we will make a connection with the popular case of magnetism in SU(2) Hubbard models. We will first discuss the most general case, in which the three local eigenvalues obtained by diagonalizing the AGF matrix are non degenerate, 
 while in a second step we present results for the case in which two levels remain degenerate. 

\subsection{Density-driven Mott transition for non-degenerate levels}
In this section we present our DMFT results for the case $\tau_{13} = 0$ and $\tau_{12} = \tau_{23} \equiv \tau$, corresponding to nearest-neighbor hopping with open boundary conditions in the synthetic dimension. With this choice, the eigenvalues of the AGF Hamiltonian are $\{\lambda_a\} = \left\{-\sqrt{2}\,\tau,\,0,\,\sqrt{2}\,\tau\right\}$.   We will argue that the numerical values of the eigenenergies do not strongly influence the physics and that the results of this section are representative of generic parameters leading to a non-degenerate local spectrum. 

We can restrict ourselves to $\mu > 0$, which corresponds to a filling ranging from half- ($n=3/2$) to full-filled bands ($n=3$). 
The information about negative values of $\mu$ and densities lower than half-filling is obtained exploiting the symmetry under a particle-hole transformation $\cd_{\bR a}\to (-1)^R\cc_{\bR a}$ plus an exchange of the flavor indices $1\leftrightarrow 3$ which transforms the Hamiltonian Eq. (\ref{Hub:Ham:1}) into itself but with $\mu\to-\mu$. 
\begin{figure}
\includegraphics[width = \columnwidth]{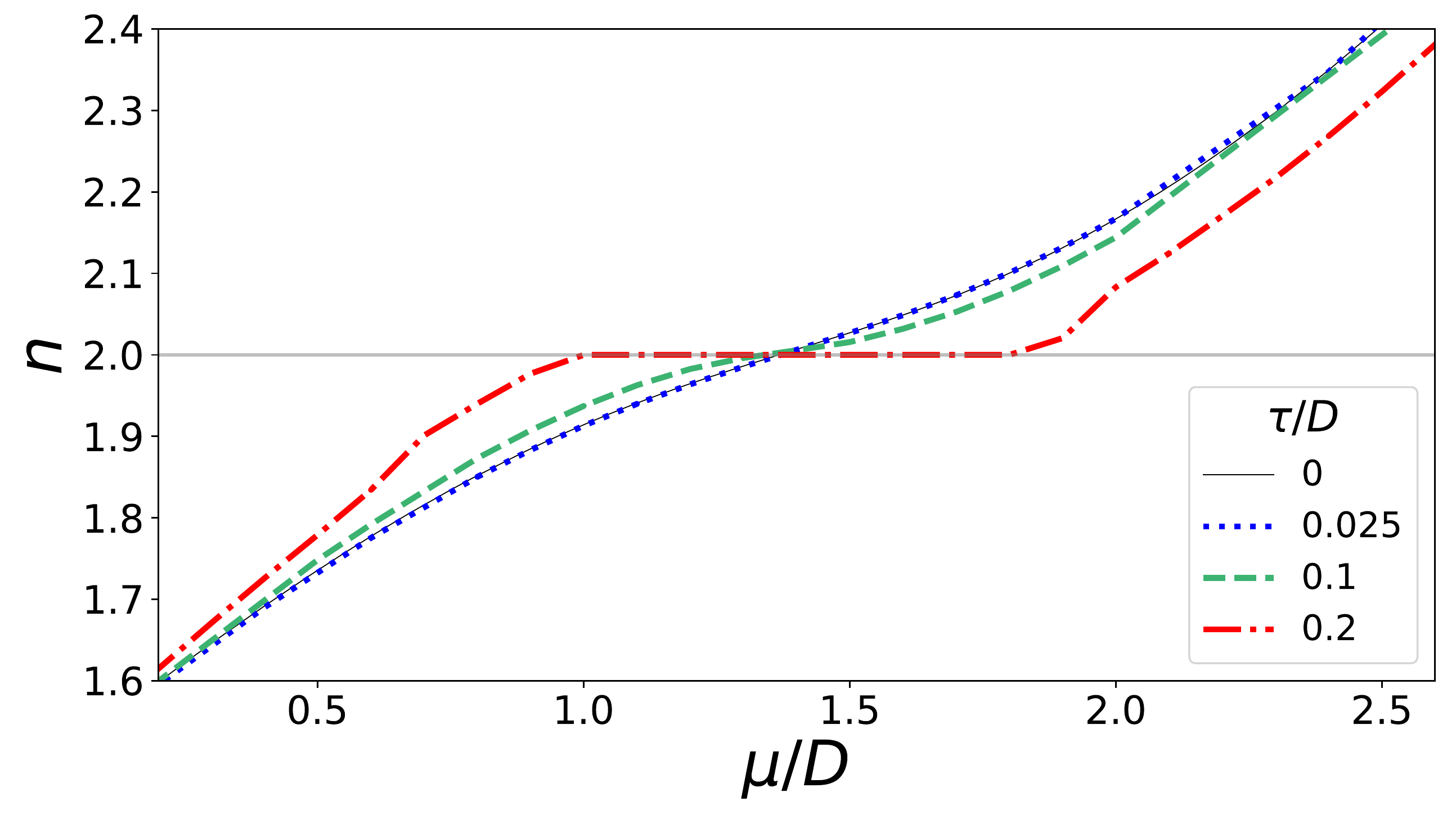}
\caption{Total density as a function of the chemical potential for several values of $\tau/D$ and $U/D = 2.5$.}
\label{U25_dens}
\end{figure}
In the absence of AGF, the three levels are degenerate and a Mott transition occurs for $n=2$ at the critical value $U_c\sim 3.5\, D$. A simple way to visualize a metal-insulator transition and the stability of a correlated system is to study the evolution of the density as a function of the chemical potential $\mu$. The derivative of this curve is the charge compressibility $\kappa = \partial n/\partial\mu$. A vanishing compressibility corresponds to a gapped state, while a divergent $\kappa$ signals an instability of the system towards phase separation.  At the same transtion the effective mass $m^*$ (not shown) diverges signaling that the particles have been localized by the mutual repulsion. 

 In Fig. \ref{U25_dens} we show the results for $U/D = 2.5$. In the absence of AGF we are below the Mott transition, as shown by the $\tau=0$ curve, which smoothly grows as a function of $\mu$ as for a metal. As $\tau$ is increased  the $n(\mu)$ curves do not change much until a critical value is reached, where a plateau appears signaling the opening of a Mott gap. We are therefore in a region of interactions where the AGF turns a metal into a correlation-driven insulator.  This result may appear counterintuitive, since the AGF represents a hopping term in the synthetic dimension which might be expected to favor delocalization and metallic states.
 \begin{figure}
\includegraphics[width = \columnwidth]{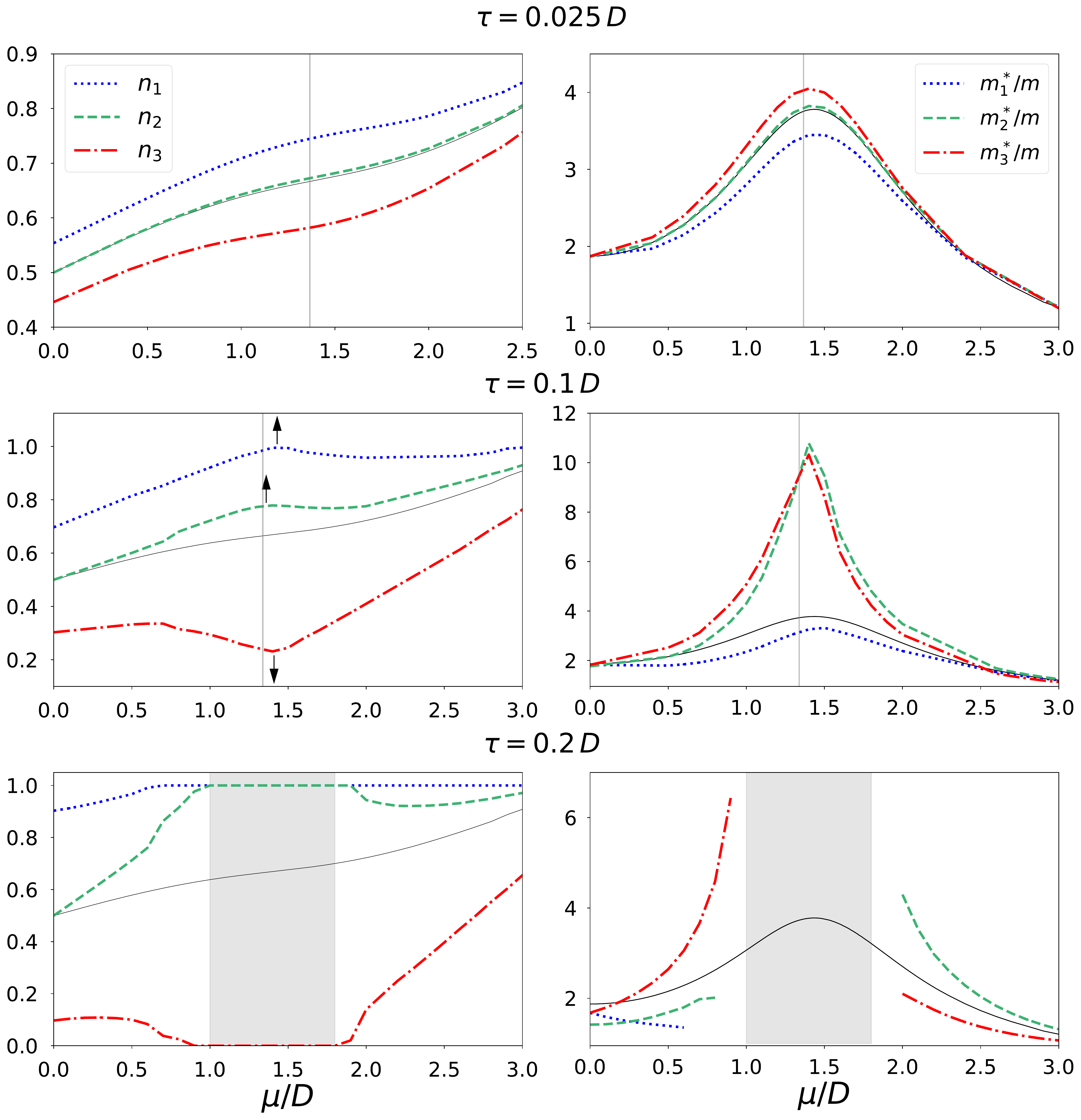}
\caption{\emph{(Left Column)} Occupation numbers of the three fermionic components  $a = 1,2,3$  as functions of the chemical potential for several values of $\tau$ and for $U/D = 2.5$. \emph{(Right Column)} Effective masses of the three fermionic components  $a = 1,2,3$  as functions of the  chemical potential for several values of $\tau$ and for $U/D = 2.5$. }
\label{U25_occs}
\end{figure}

We now show that this effect can be understood in terms of a correlation-induced redistribution of weight between the different local levels. In the left panels  of Fig. \ref{U25_occs} we show the occupation numbers of the different fermionic species $n_a \equiv \frac{1}{N_s}\sum_\bR \av{\cd_{\bR a} \cc_{\bR  a}}$ as a function of the chemical potential for fixed $\tau$.  Of course the broken degeneracy implies that the three components will have different occupation already for the non-interacting system, but we will show that this unbalance is strongly emphasized by strong correlations, in particular in the regions where the density approaches the commensurate value $n=2$ where Mott localization is possible.

For the smallest value $\tau/D = 0.025$ (first row of Fig. \ref{U25_occs}) the combined effect of $\tau$ and the interactions is only a quantitative deviation of the occupation numbers with respect to their values in the symmetric case at $\tau = 0$. Obviously the flavor with lower energy (labeled as "1") is more populated and the one with higher energy ("3")  is less populated, but all the curves are monotonically increasing. No special behavior is found when the curves cross  the commensurate filling $n=2$ (marked by a vertical line) where a Mott transition would be possible. In the right panels we show the flavor-resolved effective mass renormalizations $m_a^*/m$ (where $m$ is the bare mass of the lattice fermions in the absence of the interactions), that measures the reduction of the coherent motion of the fermions due to the interactions and it is the most direct measure of Mott localization. On the other hand, the effective mass is not easily accessible in experiments, where related quantities, like the evolution of double occupations can be measured more easily. However, since the aim of this work is to establish the theoretical possibility of selective regimes, we prefer to study the effective mass at the price of a weaker connection with experiments.

Here we observe that $m_a^*/m$ has a maximum around $n=2$ (vertical line) and that it acquires a mild dependence on the flavor, but it does not diverge, as it would happen at a standard Mott transition. Notice that the largest effective mass is found for the least populated orbital,  whose density becomes close to half-filling (0.5) when the whole system is close to $n=2$.

When $\tau$ is increased (see the data for $\tau=0.1\,D$ in the second row of Fig. \ref{U25_occs}), the spread between the densities is much larger despite the Hubbard repulsion has not been changed. This already shows stronger correlation effects induced by the AGF. Increasing the chemical potential  from $0$ to $\sim 1.4\,D$ leads to an almost complete polarization of  the majority component "1" whose occupation reaches almost the saturation value of $n_1=1$ when the system as a whole reaches a filling of $n=2$. Also the component "2" grows, while the third state is significantly depleted by increasing $\mu$. These effects reflect the large spin susceptibilities that we discussed in the previous subsection.

However the two minority components overall have an occupation of $\sim 1$ which corresponds to something similar to a half-filled $SU(2)$ Hubbard model. The system behaves therefore as the superposition of a trivial fully polarized spin species and a polarized strongly correlated metal with a sizeable spin polarization. The system remains metallic, but the two minority components are very close to Mott localization as shown by the very large effective masses, while the polarized component has only a small effective mass enhancement. If the analogy with the polarized $SU(2)$ Hubbard models holds completely, one would expect a first-order transition to a Mott insulator by further increasing $U$, as we discuss in the following. 

We finally spend a word about the negative derivative of $n_a$ with respect to $\mu$ in some regions, namely of $n_3$ when the global density is just below 2 and for the other two species for $n$ slightly larger than 2. These negative slopes emerge as a consequence of the fast polarization characteristic of our correlated system. 
Indeed, this effect is maximized around $n = 2$ (when Mott Localization is possible), where the density profiles are somehow "stretched" by the correlations, feature that we highlight drawing arrows in Fig. \ref{U25_occs}.
Of course such negative values do not correspond to an actual instability because the full compressibility is positive, but they reflect the extreme sensitivity of the response to the AGF.

\begin{figure}
\includegraphics[width = 1.05\columnwidth]{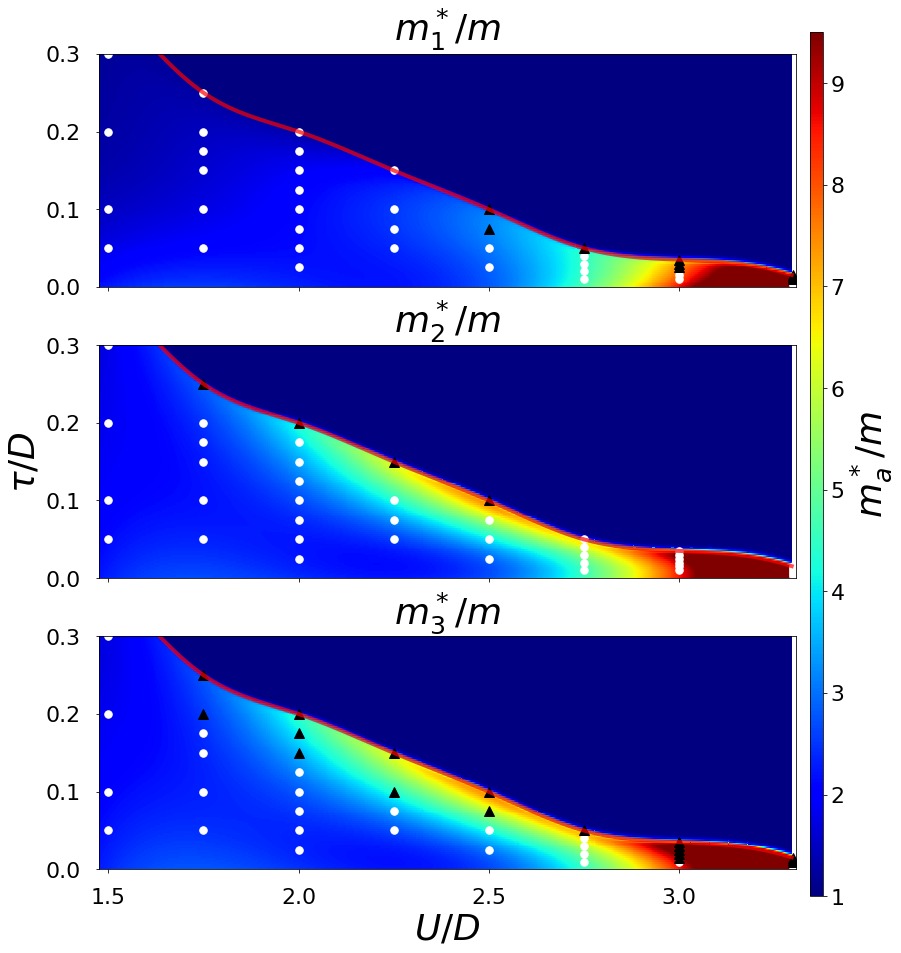}
\caption{Color plot of the effective masses of the three fermionic components evaluated at $n = 2$ on the plane $(U,\tau)$. Symbols are drawn over the density plots, indicating whether the occupation number relative to the $a$-th species has a monotonic (circles) or non-monotonic (triangles) trend as a function of $\mu$. }
\label{density_plot_masses}
\end{figure}

At $\tau/D = 0.2$  the $n$-$\mu$ curve becomes flat in a region of $\mu$ signalling an incompressible insulating phase (see Fig. \ref{U25_dens}). The trends of the populations as a function of $\mu$ are even more pronounced than in the previous case. As shown in the lower row of Fig. \ref{U25_occs}, in the whole window where $n=2$ the bands relative to the flavors $a = 1,2$ are totally occupied ($n_{1/2} = 1$) while the band of the $a = 3$ species is empty ($n_3 = 0$). We have therefore a kind of band insulator, which, however, would not be stabilized in the absence of interactions which enhance the spin polarization with respect to the non-interacting case. 
The non-trivial nature of this state is confirmed by the non-monotonic behavior of $n_a(\mu)$ in proximity of the transition and by the evolution of $m_a^*/m$ . 
All the effective masses are smaller in magnitude with respect to $\tau/D = 0.1$, but they are enhanced when the $n=2$ region is approached from both sides (larger and smaller densities). Interestingly the mass enhancement for each component is more pronounced in the regions where the "flavor compressibility" is negative and this leads to a reversal of the hierarchy of the effective masses on the two sides of the transition. 

\subsection{The $n=2$ phase diagram}

In Fig. \ref{density_plot_masses}  we summarize our results for the flavor-selective correlation effects in a sort of phase diagram where we report  color plots of the effective masses of the three fermionic species in the $(U,\tau)$ plane in three separated plots which help to highlight at the same time the common trends between all the species and their differences. 

 We sit at the characteristic density $n = 2$ where correlation-driven insulating phases are possible and all the correlations effects are more visible. We superimpose black triangles to mark the points where  the  occupation of each species is non-monotonic, while circles indicate a regular behavior. The correlation between the non-monotonicity and a large $m_a^*$ is clear and completely generic.

The solid red line marks the critical interaction strength for the metal insulator transition. As expected, for $\tau =0$ we recover the standard second-order Mott transition for the SU(3) model and a critical  $U \sim 3.5 D$. As $\tau$ increases the transition occurs for smaller $U_c$ and it becomes of first-order, analogously to the standard SU(2) Hubbard model in a magnetic field. 
If we move along the $U$-axis at fixed $\tau$ we always find an increase of the three effective masses, while a richer behavior is found as a function of $\tau$ and fixed $U$. For every value of the interaction $m^*_1$ decreases along the $\tau$-axis while $m_{2/3}^*$ increases. 

The reason for this strong differentiation is the one we described above. The main driving force behind this phase diagram is indeed the rapid enhancement of the unbalance between the different species dictated by AGF splitting.
Even small deviations from $SU(3)$ symmetry are strongly emphasized by correlation effects, leading to a strong polarization and a strong dependence of the correlation properties on the flavor.

In fact, the correlations-enhanced susceptibility  leads to an almost full polarization of the flavor with the lowest on-site energy. This spin species behaves as a metal close to a transition toward a band-insulator, whose correlations are negligible.
This is confirmed by the behavior of $m^*_1$ that decreases smoothly as a function of the external perturbation.

Therefore  the formation of local moments, and the increase of effective mass, both associated with Mott physics, only involve the two remaining flavors.
In this way the system can gain energy both from the polarization of the flavor with the lowest on-site energy and the Mott localization involving the two other flavors with higher on-site energies.

 This effect does not depend strongly on the  precise values of the bare splittings and we can obtain very similar results in cases where the splittings are not equally spaced as in our example. For this reason the present results are representative of arbitrary situations where the three flavors are not degenerate. 

The strong correlation-induced enhancement of the bare splittings is reminiscent of the renormalization of fhe crystal-field splittings found in multi-orbital SU(2) Hubbard models which can trigger strong orbital selectivity or even orbital-selective Mott transitions. \cite{PhysRevB.76.085127, PhysRevB.87.205108, PhysRevLett.117.176401}

Indeed the trend of  $m^*_{2,3}$ is reminiscent also of the behavior of the fermions in the single band Hubbard model where a dramatic enhancement of the magnetization as a function of the intensity of an external magnetic field is observed in the intermediate coupling regime\cite{bauer2007field,bauer2009quasiparticle,laloux1994effect,held1997correlated}.
 Here, a metamagnetic behavior has been reported with a positive derivative of the finite-field magnetic susceptibility which can be associated to a   strong renormalization of the masses which occurs just before a polarized phase is reached \cite{bauer2007field,bauer2009quasiparticle}. For this reason we label this regime as a \emph{selective metamagnetic} behavior, where a fully polarized species coexists with a metamagnetic pair of flavours. 
We stress again that the regions with a large magnetic response coincide with negative "flavor compressibility"   $\kappa_a \equiv \frac{\partial n_a}{\partial \mu}$ for some species. While this property does not imply an actual thermodynamic instability as the actual compressibility is positive, it may be used to devise experimental protocols where a stable system is turned into an unstable one by the sudden removal of the species with positive flavor compressibility.

\subsection{Density-driven Mott transition for the doubly degenerate case}
We now consider the limiting case where two species lie on a degenerate manifold and a residual $SU(2)$ symmetry is preserved between them.  
For $\tau_{12} = \tau_{23} = \tau_{31} \equiv \tau$, which corresponds to periodic boundary conditions in the artificial dimension,
the eigenvalues of the AGF matrix read $\{\lambda_a\} =\{-2\tau,\tau,\tau\}$, with one single level at lower energy than a degenerate doublet. In contrast with the previous case, the $n=1$ and $2$ solutions cannot be transformed one into the other through a particle-hole transformation and they give rise to different transitions. If we change the sign of $\tau$, we reverse the spectrum and we exchange the nature of the two transitions as well.

For positive $\tau$ the transition the physics in the regime $n \sim 1$ is very similar to the $SU(2)$ case at half-filling in presence of a magnetic field, with the only difference that the high-energy level is now degenerate as opposed to the majority spin of the $SU(2)$ system, but this modification does not change significantly the results. We do not discuss these results as they essentially reproduce those reported, e.g., in Refs.\cite{bauer2007field,bauer2009quasiparticle} and they do not give rise to new physical regimes.

\begin{figure}[htbp!]
\includegraphics[width = \columnwidth]{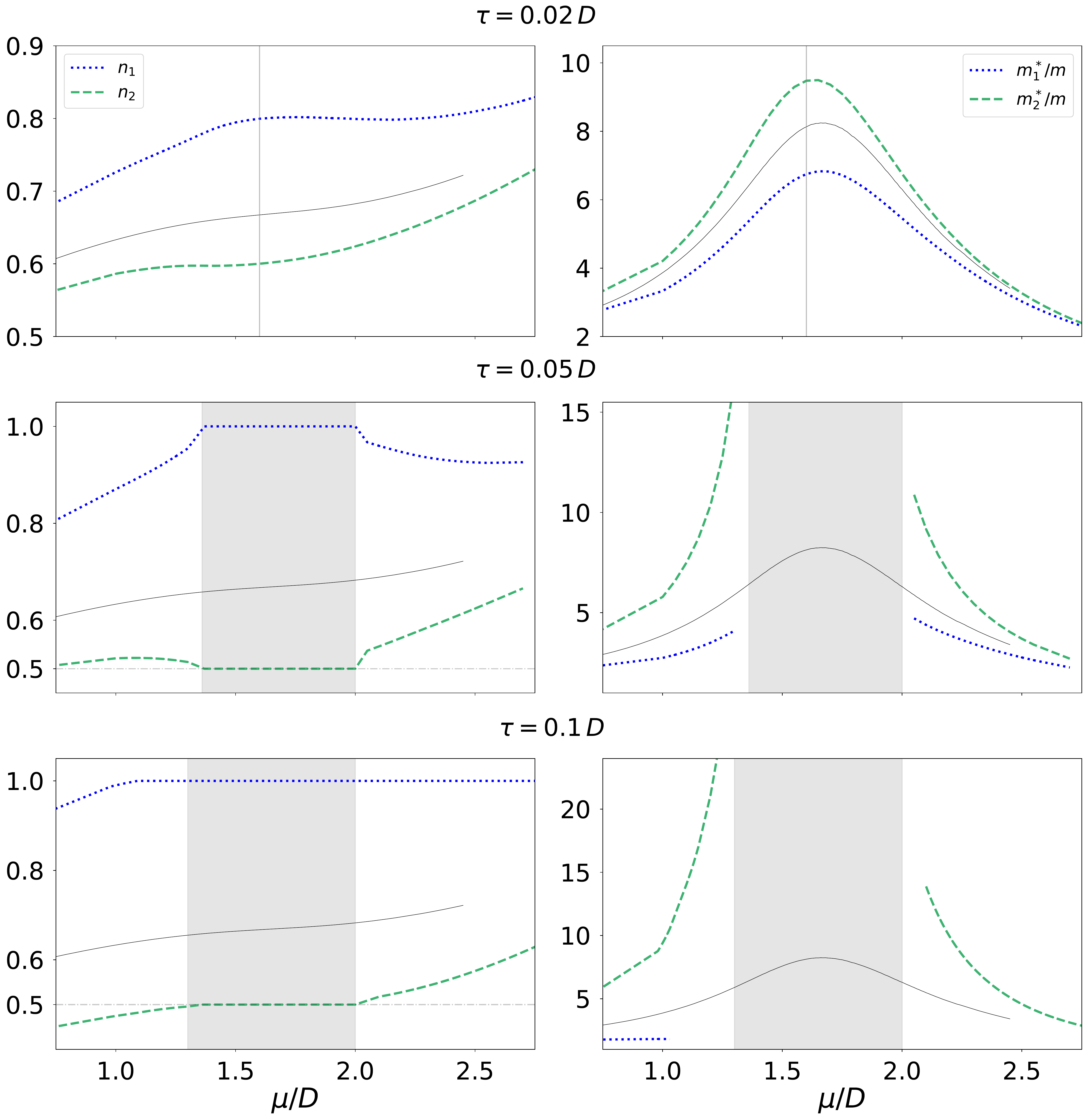}
\caption{(Left) Occupation numbers of the three fermionic components  $a = 1,2$  as function of the chemical potential for several values of $\tau$ and for $U/D = 3.0$. (Right) Renormalized masses of the three fermionic species $a=1,2$ for several values of $\tau$ and for $U/D = 3.0$.}
\label{occs_limiting_2}
\end{figure}

The $n \sim 2$ case displays instead a characteristic behavior, which can be connected with the non-degenerate case.
Fig. \ref{occs_limiting_2}  shows the occupation numbers $n_1$ and $n_2 = n_3$ as functions of the chemical potential for  $\tau/D = 0.02,\,0.05,\,0.1$. We set $U/D = 3.0$, for which the $\tau=0$ system is in an itinerant state in the symmetric case  (black solid line) and $n_a(\mu)$ are increasing monotonic functions. For $\tau/D = 0.02$ no gaps are open and the system is still metallic, but the occupation numbers start to display a non-monotonic trend and the effective masses are quite large. When $\tau/D = 0.05$  the occupation numbers display a plateau at integer filling, with the non-degenerate species becoming fully polarized and the two degenerate species with a density of 0.5 fermions per site, which is characteristic of a nonmagnetic Mott insulator for the SU(2) Hubbard model.

Therefore also in this case, we observe a metal-insulator transition driven by the AGF, but the insulating state here is a mixture between a Mott insulator composed by the two degenerate species with $n_2 = n_3 = 0.5$ and a fully polarized state with $n_1 = 1$. The anomalous non-monotonic behavior of the occupation numbers flanks the transition.
For $\tau/D = 0.1$, we still find the same insulating phase, but the occupation numbers have lost their peculiar non-monotonic trend. In fact, for very large values of $\tau$, $n_1 = 1$ regardless of the value of the chemical potential and the Mott transition of the $SU(2)$-Hubbard model is recovered.
The AGF term turns the system into a Mott insulator (plus a polarized band) because it reduces the effective degeneracy, which makes Mott localization easier because of a smaller kinetic energy to compete with the Hubbard repulsion\cite{florens2002mott}. We can thus view  $\tau$ as a field which  interpolates from the $SU(3)$-Hubbard model to the $SU(2)$-Hubbard model. We also notice that in this case the effective mass enhancements are significantly stronger than in the non-degenerate case.

\section{Conclusions}
In this manuscript we have investigated the effect of Raman processes on a three-component fermionic fluid in an optical lattice. We addressed the simple case where the fields are described by real numbers, which allows us to focus on effects related to Mott physics in the presence of a lifting of the SU(3) degeneracy. We postpone to future studies the analysis of complex fields and the related topological properties in the presence of strong correlations. The focus of this work is on the theoretical possibility of these phases and we do not address the direct experimental detection of these phenomena, which we postpone to future work where the effect of the trapping potential will also be considered.

We find a variety of novel phenomena which depend on the values of the fields and on the density of fermions. In the general case where the fields completely break the $SU(3)$ symmetry, we find that the AGF favor Mott localization and give rise to strongly flavor-selective physics where the densities and the effective masses of the different carriers have remarkably different behaviors. The most populated orbital becomes completely polarized, while the two others behave similarly to a SU(2) Hubbard model in a magnetic field with a characteristic magnetic response and large effective mass. Furthermore, we observe a non-monotonic trend of the occupations of individual species as functions of the chemical potential, which would signal a sort of "flavor-selective" phase separation even though the system as a whole is thermodynamically stable. 

When two levels are degenerate we find instead an insulating state where a polarized flavor coexists with a two-component Mott insulator. Also this transition is preceeded by a similar "flavor-selective" instability of some of the fermionic species. Our results confirm the richness of the phase diagrams of multi-component fermionic systems which is a  highly debated topic in solid-state physics.
These phenomena can be directly observed in current experimental set-ups in cold atoms, and they open a number of interesting perspectives, ranging from the effect of complex AGF to increasing the number of components. In the former case, it has been already shown that topologically protected edge states emerge in the non-interacting system. The effect of strong interactions on a topological phase transition can lead to remarkable phenomena like a first-order topological transition\cite{PhysRevLett.114.185701} or a Mott localization and successive reconstruction of edge states\cite{PhysRevB.95.205120}.
\section*{Acknowledgments}
We thank G. Cappellini, J. Catani and L. Fallani for precious discussions.  We acknowledges financial support from MIUR through the PRIN 2015 program (Prot. 2015C5SEJJ001),  SISSA/CNR project "Superconductivity, Ferroelectricity and Magnetism in bad metals"  and FSE-Friuli Venezia Giulia for the HEaD "Higher Education And Development" project FP1619889004. 
\\
\appendix
\section{Derivation of the Bethe-Salpeter equation}\label{Der:BS}
In this appendix we shall derive eq.(\ref{Bethe:Salpeter}) following mainly the spirit of Ref.\cite{tahvildar1997magnetic}. 
We start by noticing that the magnetic susceptibility $\chi^{\alpha}_\beta(\bR-\bR^\prime)\equiv \frac{\partial\langle S^{\alpha}_\bR\rangle}{\partial h^{\beta}_\bR}$ can be written as:
\begin{equation}
\chi^{\alpha}_\beta(\bR-\bR^\prime) = T\sum_\nu \left(
\frac{\partial G^\alpha_{\bR\bR}(\nu)}{\partial h^\beta_{\bR^\prime}}-\frac{\partial G^N_{\bR\bR}(\nu)}{\partial h^\beta_{\bR^\prime}}
\right)e^{-i\,\nu\,0^-},
\end{equation}
where $G^a_{\bR\bR^\prime}(\nu) \equiv- \int_0^\beta d\tau e^{i \nu \tau} T_{\tau}\left\langle\cc_{\bR a}(\tau)\cd_{\bR^\prime a}(0)\right\rangle$, $\nu = (2 n + 1)\pi\,T$ are the fermionic Matsubara frequencies, and $\alpha,\beta = 1,2,...,N-1$.
The Dyson's equation in real space of the Green's function reads:
\begin{equation}
G^{-1}_{\bR\bR^\prime}(\nu)  = \left[G_0^{a}\right]^{-1}_{\bR\bR^\prime}(\nu) + \lambda^a_{\bR}\delta_{\bR\bR^\prime} - \Sigma_{\bR\bR^\prime}^a(\nu),
\end{equation}
where $G_0$ is the non-interacting Green's function in absence of the magnetic field, $\Sigma$ is the self-energy of the system.
Therefore $G$ is an $N_s\times N_s$ matrix whose indices are given by $\bR, \bR^\prime$, and if we use the matrix identity $\partial_{x}(G) = -G\partial_x(G^{-1})G$, where $x$ is some parameter upon which the matrix G depends, we have the following identity:
\begin{widetext}
\begin{equation}\label{Bethe:Real:Space}
\frac{\partial G^a_{\bR\bR}}{\partial h ^\beta_{\bR^\prime}} =-(\delta_{a\,\beta}-\delta_{a\, N})G_{\bR\bR^\prime}(\nu)G_{\bR^\prime\bR}(\nu) +\sum_{\bR_1 \bR_2 \bR_1^\prime \bR_2^\prime}\sum_{b\nu^\prime}G_{\bR\bR_1}(\nu)G_{\bR_2 \bR}(\nu)\,\Gamma_{\bR_1 \bR_2;\bR_1^\prime \bR_2^\prime}^{a\,b}(\nu,\nu^\prime)\,\frac{\partial G_{\bR_1^\prime \bR_2^\prime}}{\partial h^{\beta}_{\bR^\prime}}(\nu^\prime),
\end{equation}
\end{widetext}
where $\Gamma_{\bR_1\bR_2;\bR_1^\prime \bR_2^\prime}^{ab}(\nu,\nu^\prime) \equiv \frac{\partial \Sigma^{a}_{\bR_1\bR_2}(\nu)}{\partial G^b_{\bR_1^\prime \bR_2^\prime}(\nu^\prime)}$ is 
the two-particle irreducible vertex function. In the limit of infinite dimensions we can keep only the local contributions of the vertex function in eq.(\ref{Bethe:Real:Space}), i.e. $\Gamma_{\bR_1\bR_2;\bR_1^\prime \bR_2^\prime}^{ab}(\nu,\nu^\prime) \to \Gamma_{loc}^{ab}(\nu,\nu^\prime)\delta_{\bR_1\bR_2}\delta_{\bR_1^\prime \bR_2^\prime}\delta_{\bR_1\bR_1^\prime}$. After expanding all the functions of the lattice sites in their Fourier components we obtain:
\begin{widetext}
\begin{equation}
\widetilde{\chi}^{a}_\beta(\nu,\bq) = (\delta_{a \,\beta}-\delta_{a\,N})\chi_0(\nu,\bq)-\chi_{0}(\nu,\bq)\sum_{b\nu^\prime}\Gamma^{ab}_{loc}(\nu,\nu^\prime)\widetilde{\chi}^{b}_{\beta}(\nu^\prime,\bq),
\end{equation}
\end{widetext}
where $\widetilde{\chi}^{a}_\beta(\nu,\bq) \equiv \sum_{\bR}\exp[-i\,(\bR-\bR^\prime)\cdot\bq]\frac{\partial G^a_{\bR\bR(\nu)}}{\partial h_{\bR^\prime\beta}}$, $\chi^0(\bq,\nu)\equiv -\frac{1}{V}\sum_{\bk}G(\bk,\nu)G(\bk+\bq,\nu)$, with $G(\bk,\nu)$ being the Fourier components of the Green's function.
For symmetry reason, since all derivatives are calculated at zero field, $\Gamma^{ab} = \Gamma_{\parallel}\delta_{ab} + \Gamma_{\perp}(1-\delta_{ab})$ and therefore we obtain the following formula for $\chi^{\alpha}_\beta(\nu,\bq) \equiv \widetilde{\chi}^{\alpha}_\beta(\nu,\bq) - \widetilde{\chi}^{N}_\beta(\nu,\bq)$  with $\alpha  < N$:
\begin{widetext}
\begin{equation}\label{summed:BS}
\chi^{\alpha}_\beta(\nu,\bq) = (1 + \delta_{\alpha \beta})\chi_0(\nu,\bq)-\chi_{0}(\nu,\bq)\sum_{\nu^\prime}\Gamma_{A}(\nu,\nu^\prime) \chi^{\alpha}_\beta(\nu^\prime,\bq),
\end{equation}
\end{widetext}
where $\Gamma_A = \Gamma_\parallel - \Gamma_\perp$.

At this stage, we can define the static generalized susceptibility matrix:
\begin{widetext}
\begin{equation}
\left[\chi^{\alpha}_\beta(\bR-\bR^\prime)\right]_{\nu\nu^\prime} \equiv T \sum_{ab}c^a_{\alpha}c^b_{\beta}\int_0^\beta \prod_{i = 1}^4 d\tau_i \, T_\tau\left<\cd_{\bR a}(\tau_1)\,\cc_{\bR a}(\tau_2)\,\cd_{\bR^\prime b}(\tau_3)\cc_{\bR^\prime  b}(\tau_4)\right> e^{-i[\nu (\tau_1-\tau_2) + \nu^\prime(\tau_3-\tau_4)] },
\end{equation}
\end{widetext}
where $c^a_\alpha = \delta_{a\alpha}-\delta_{aN}$. 
The generalized susceptibility matrix is related to the thermodynamical susceptibility via the following relation: 
\begin{eqnarray}
 \chi^{\alpha}_\beta(\bR-\bR^\prime)  &=&  T\,\int_0^\beta d\tau \int_0^\beta d\tau^\prime\, T_\tau\left<\hat{S}^{\alpha}_\bR(\tau)\hat{S}^{\beta}_{\bR^\prime}(\tau^\prime)\right> \nonumber \\
 &=& T^2\sum_{\nu\nu^\prime}e^{-i \nu 0^-}e^{-i\nu^\prime 0^-}\left[\chi^{\alpha}_\beta(\bR-\bR^\prime)\right]_{\nu \nu^\prime}, \nonumber \\
\end{eqnarray}
and its Fourier components are related to the quantity $\chi^\alpha_\beta(\nu,\bq)$ in Eq.(\ref{summed:BS}) in the following way $\chi^\alpha_\beta(\nu,\bq)=T\sum_{\nu^\prime}[\chi^\alpha_\beta(\bq)]_{\nu\nu^\prime}$. 
If we define $\left[\chi_0(\bq)\right]_{\nu\nu^\prime} \equiv \beta \,\chi_0(\nu,\bq)\,\delta_{\nu\nu^\prime}$, we can write the following matrix identity for the generalized susceptibility:
\begin{equation}\label{BS}
\chi^{\alpha}_\beta(\bq) = (1+ \delta_{\alpha \beta})\,\chi_0(\bq) - T\,\chi_0(\bq)\,\Gamma_A\,\chi^{\alpha}_\beta(\bq),
\end{equation}

where, $[\chi^{\alpha\beta}(\bq)]_{\nu\nu^\prime} = \sum_\bR e^{ -i(\bR-\bR^\prime)\cdot \bq }\,\left[\chi^{\alpha\beta}_{\bR\bR^\prime}\right]_{\nu \nu^\prime}$. 
We notice that in Eq.(\ref{BS}) the symbols $\chi^\alpha_\beta(\bq)$, $\chi_0(\bq)$ and $\Gamma_A$ represent square matrices in the space of Matsubara frequencies $\nu$ and $\nu^\prime$.  
\bibliography{biblio}

\end{document}